\documentclass[aps,prd,twocolumn,superscriptaddress,10pt,nofootinbib,floatfix]{revtex4-2}
\usepackage{eurosym}
\usepackage{amssymb}
\usepackage{textcomp}
\usepackage{graphicx}
\usepackage{epsfig}
\usepackage{subfigure}
\usepackage{amsmath}
\usepackage{slashed}
\usepackage{units}
\usepackage{hyperref}
\usepackage{array}
\usepackage{lmodern}
\usepackage{dcolumn}
\usepackage{bm}

\begin{document}

\title{Violation of C/CP Symmetry Induced by a Scalar Field Emerging\\
from a Two-Brane Universe: A Gateway to Baryogenesis}
\author{Micha\"{e}l Sarrazin}
\email{michael.sarrazin@ac-besancon.fr}
\affiliation{Universit\'{e}
de Franche-Comt\'{e}, CNRS, Institut UTINAM, \'{E}quipe de Physique Th\'{e}orique, F-25000 Besan\c con, France} 
\affiliation{Laboratory of Analysis by Nuclear Reactions, Department of Physics,
University of Namur, 61 rue de Bruxelles, B-5000 Namur, Belgium}
\author{Coraline Stasser}
\affiliation{Laboratory of Analysis by Nuclear Reactions, Department of Physics,
University of Namur, 61 rue de Bruxelles, B-5000 Namur, Belgium}

\begin{abstract}
A model of baryogenesis is introduced where our usual visible Universe is a
3-brane coevolving with a hidden 3-brane in a multidimensional bulk. The
visible matter and antimatter sectors are naturally coupled with the hidden
matter and antimatter sectors, breaking the C/CP invariance and leading to
baryogenesis occurring after the quark-gluon era. The issue of leptogenesis
is also discussed. The symmetry breaking spontaneously occurs due to
the presence of an extra scalar field supported by the $U(1)\otimes U(1)$
gauge group, which extends the conventional electromagnetic gauge field in
the two-brane universe. Observational consequences are discussed.
\end{abstract}

\maketitle

\section{Introduction}

\label{Intro}

While the standard model of particle physics and the concordance model of
cosmology have achieved predictive success, there are still puzzling data
that require interpretation. These include for instance the observations of
dark matter and dark energy \cite{DM}, as well as the matter-antimatter
asymmetry \cite{baryoR,Bolt,rev1}. Our Universe is mainly empty space, with
a mean baryonic matter density about one proton per 4 cubic meters. However,
such a value is extremely large and the absence of antimatter raises
significant questions. Indeed, shortly after the initial moment of the Big
Bang, particles and antiparticles should have been in thermal equilibrium
with the photon bath. As the Universe expanded, matter and antimatter should
have almost completely annihilated once the global temperature dropped below
the mass energy of each particle. Nevertheless, a large baryon-antibaryon
asymmetry is observed, with the visible Universe today dominated by matter
rather than antimatter \cite{baryoR,Bolt,rev1}. This is the baryogenesis
problem. Those unresolved issues, coupled with the quest for a unified
theory of fundamental interactions, have motivated extensive theoretical
work, resulting in a diverse landscape of models that challenge new
experimental projects aimed at testing new physics \cite%
{DM,baryoR,Models,newXp,BranR,mirr}. In this context, many theoretical works
suggest that our visible Universe could be a 3-dimensional physical entity
(a $3-$brane) embedded in a $(3+N,1)$ $- $space-time ($N\geq 1$) known as
the bulk \cite{br1,br2,br3,br4,br5,br6,br7}. Hidden $3-$branes may coexist
alongside our own in the bulk. This leads to a rich phenomenology
encompassing both particle physics and cosmology \cite{BranR}. Some studies
propose that hidden branes could host dark matter, or that interactions
between branes could account for dark energy \cite{RevBr1, RevBr2, DMBrane1,
DMBrane2, Phant1, Phant2}. In addition, many scenarios suggest that the Big
Bang was triggered by a collision between our visible brane and a hidden one 
\cite{Branerev,cob1,cob2,cob3,cob4,cob5,cob6,cob7,cob8,cob9,cob10,cob11}.
Previous research has highlighted that braneworld scenarios or dark matter
models involving sterile particles could explain baryogenesis \cite{Dvali,
Ber}.

Moreover, numerous theoretical predictions have emerged regarding hidden or
dark sectors, allowing phenomena like neutron--hidden neutron transitions $%
n-n^{\prime }$ \cite{mirr,pheno}. Over the past decade, this phenomenology
has prompted efforts to constrain these scenarios through neutron
disappearance/reappearance experiments \cite%
{Exp0,Exp1,Exp2,Exp3,Exp4,Exp5,Exp6}. Specifically, a neutron $n$ in our
visible brane can transmute into a hidden neutron $n^{\prime }$, effectively
swapping into a hidden brane \cite{PLB,M4xZ2,cp1,cp2}, depending on a
specific coupling constant $g$ between visible and hidden sectors. The
theoretical study of this brane phenomenology \cite{PLB,M4xZ2,cp1,cp2} has
been complemented by experimental tests over the past two decades \cite%
{Exp0,Exp1,Exp2,Exp3,Exp4}, particularly through passing-through-wall
neutron experiments \cite{Exp1,Exp2}, which have provided stringent bounds
on the coupling constant $g$ \cite{Exp3,Exp4}.

In the present paper, assuming previous theoretical results \cite%
{M4xZ2,cp1,cp2}, one shows how a two-brane universe provides a solution to
the baryogenesis issue after the phase transition from quark-gluon plasma to
hadron gas. In particular, the violation of the C/CP symmetry naturally
arises in the two-brane universe model through the occurrence of a scalar
field resulting from the splitting of the electromagnetic gauge field on
each brane. Due to the scalar field, a dressed coupling constant $\mathfrak{g%
}$ then replaces the bare coupling constant $g$. The coupling constant $%
\overline{\mathfrak{g}}$ describing the $\overline{n}-\overline{n}^{\prime }$
transition between the antineutron and hidden antineutron sectors then
differs from $\mathfrak{g}$. Consequently, $\overline{n}-\overline{n}%
^{\prime }$ transitions would occur at a different rate than $n-n^{\prime }$
transitions with an asymmetry allowing the current baryon-antibaryon ratio
with respect of the Sakharov conditions \cite{baryoR,Cond}.

The study is organized as follows. In Section II, one provides a brief
overview of the theoretical framework used here and previously introduced in
literature \cite{PLB,pheno,M4xZ2,cp1,cp2}, and which enables the study of
particle dynamics in a two-brane universe. In Section III, one shows how the
electromagnetic gauge field $U(1)\otimes U(1)$ in a two-brane universe
naturally replaces the $U(1)$ gauge field, and how an additional pseudo
scalar field then arises. The properties of the vacuum state and of the
fluctuations of this new field are clarified in section IV. One then shows
and discusses how this field breaks the C/CP symmetry in Section V, also
introducing the interbrane coupling Hamiltonian. Next, in Section VI, it is
shown that the coupling constant $\overline{\mathfrak{g}}$ between the
antineutron and hidden antineutron sectors must then differ from $\mathfrak{g%
}$. Both coupling constants $\mathfrak{g}$ and $\overline{\mathfrak{g}}$ are
naturally affected by the scalar field, leading to the expected conditions
for baryogenesis. In section VII, from the interbrane coupling Hamiltonian,
one introduces the Boltzmann equations relevant to describe the baryogenesis
in a two-brane universe. Finally, before concluding, the results obtained
from these equations are shown and discussed in the section VIII. One shows
thus the relevance of the mechanism inducing the C/CP violation to explain
baryogenesis in the context of braneworld scenarios. One also discusses the
ways to observationally constrain the present baryogenesis model.

\section{Theoretical framework of the fermion dynamics in a two-brane
universe}

\label{Theo}

Braneworld physics and cosmology can present a complex landscape of models,
making their study challenging. However, over the past two decades, it has
been shown \cite{PLB,M4xZ2,cp1,cp2} that this study can be simplified
through a mathematical and physical equivalence between two-brane universes
and noncommutative two-sheeted spacetimes. The reader is encouraged to
consult the cited references \cite{M4xZ2,cp1,cp2,dem} for the demonstrations
of this equivalence not depicted here, for the sake of clarity.

To be more precise, let us consider a two-brane universe in a $(3+N,1)$ $-$%
bulk ($N\geq 1$). Each brane has a thickness $M_{B}^{-1}$ along extra
dimensions - with $M_{B}$ the brane energy scale - and $d$ is the distance
between both branes in the bulk. Then, at the sub-GeV-scale, the quantum
dynamics of fermions in the two-brane universe is the same as in a
two-sheeted space-time $M_{4}$\ $\times $\ $Z_{2}$\ described with
noncommutative geometry \cite{M4xZ2,cp1,cp2,dem}.

The phenomenological discrete space-time $M_{4}$ $\times $ $Z_{2}$ replaces
the physical continuous $(3+N,1)$ $-$bulk ($N\geq 1$) with its two branes 
\cite{M4xZ2,cp1,cp2}. At each point along the discrete extra dimension $%
Z_{2} $, there is a four-dimensional space-time $M_{4}$ endowed with its own
metric. Each $M_{4}$ sheet describes each braneworld considered as being
separated by a phenomenological distance $\delta =1/g$, with $g$ the bare
coupling constant between fermionic sectors. $g$ is a function against $%
M_{B} $, $d$ and also the mass of the fermion under consideration \cite%
{M4xZ2,cp1,cp2}. The function can also depend on the bulk properties (i.e.
dimensionality and compactification). For instance, for neutron and a $%
M_{4}\times R_{1}$ bulk, one gets \cite{cp1,cp2}: 
\begin{equation}
g\sim \frac{m_{Q}^{2}}{M_{B}}e^{-m_{Q}d},  \label{gf}
\end{equation}%
where $m_{Q}$ is the mass of the quark constituents in the neutron -- i.e.
the mass of the quarks up and down dressed with gluons fields and virtual
quarks fields such that $m_{Q}=m_{up}=m_{down}=327$ MeV \cite%
{CQMbook,CQM1,CQM2,CQM3}.

The effective $M_{4}\times Z_{2}$ Lagrangian for the fermion dynamics in a
two-brane Universe is \cite{M4xZ2,cp1,cp2}: 
\begin{equation}
\mathcal{L}_{M_{4}\times Z_{2}}\sim \overline{\Psi }\left( {i{\slashed{D}}-m}%
\right) \Psi .  \label{LM4Z2}
\end{equation}%
Labeling $(+)$ (respectively $(-)$) our brane (respectively the hidden
brane), one writes: $\Psi =\left( 
\begin{array}{c}
\psi _{+} \\ 
\psi _{-}%
\end{array}%
\right) $ where $\psi _{\pm }$ are the wave functions in the branes $(\pm )$
and $m$ is the mass of the bound fermion on a brane, here the quark
constituent. The derivative operators acting on $M_{4}$ and $Z_{2}$ are $%
D_{\mu }=\mathbf{1}_{8\times 8}\partial _{\mu }$ ($\mu =0,1,2,3$) and$\
D_{5}=ig\sigma _{2}\otimes \mathbf{1}_{4\times 4}$, respectively, and the
Dirac operator acting on $M_{4}\times Z_{2}$ is defined as ${\slashed{D}=}%
\Gamma ^{N}D_{N}=\Gamma ^{\mu }D_{\mu }+\Gamma ^{5}D_{5}$ where: $\Gamma
^{\mu }=\mathbf{1}_{2\times 2}\otimes \gamma ^{\mu }$\ and\ $\Gamma
^{5}=\sigma _{3}\otimes \gamma ^{5}$. $\gamma ^{\mu }$ and $\gamma
^{5}=i\gamma ^{0}\gamma ^{1}\gamma ^{2}\gamma ^{3}$ are the usual Dirac
matrices and $\sigma _{k}$ ($k=1,2,3$) the Pauli matrices. Eq. (\ref{LM4Z2})
is characteristic of fermions in noncommutative $M_{4}\times Z_{2}$
two-sheeted space-times as introduced by other authors \cite%
{NC1,NC2,NC3,NC4,NC5,NC6,NC7,NC8}.

One refers to the terms proportional to $g$ as geometrical mixing \cite%
{M4xZ2,cp1,cp2}. The present approach serves as a valuable tool for
investigating the phenomenology of braneworlds and exploring their
implications within realistic experimental settings \cite%
{Exp0,Exp1,Exp2,Exp3,Exp4}.

In the following sections, one shows how the violation of C/CP symmetry
naturally arises from the $M_{4} \times Z_{2}$ framework, using the scalar
field that emerges from the splitting of the electromagnetic gauge field.
Therefore, it is necessary to consider $U(1)\otimes U(1)$ instead of $U(1)$.

\section{Gauge field and extra scalar field}

\label{Gauge}

In a two-brane universe, the electromagnetic field is described by the
effective $U(1)_{+}\otimes U(1)_{-}$ gauge field in the $M_{4}\times Z_{2}$
space-time \cite{M4xZ2}. Here, $U(1)_{+}$ is the gauge group associated with
the photon field localized on our brane, while $U(1)_{-}$ is the gauge group
of the photon field localized on the hidden brane. This is not merely a
corollary of the $M_{4}\times Z_{2}$ description, but a demonstrated
consequence when examining the low-energy dynamics of fermions in the
two-brane system\footnote{%
It is noteworthy that the phenomenology of the gauge group $U(1) \otimes
U(1) $ also manifests in other contexts beyond brane physics \cite%
{2G1,2G2,NC1,NC2,NC3,NC4,NC5,NC6,NC7,NC8}.}\cite{M4xZ2}. The group
representation is therefore: 
\begin{equation}
G=\text{diag}\left\{ \exp (-iq\Lambda _{+}),\exp (-iq\Lambda _{-})\right\} .
\label{GG}
\end{equation}%
Looking for an appropriate gauge field such that the gauge covariant
derivative is ${\slashed{D}}_{A}\rightarrow {\slashed{D}}+iq\slashed{A}$
with the following gauge transformation rule:%
\begin{equation}
\slashed{A}^{\prime }=G\slashed{A}G^{\dagger }-\frac{i}{q}G\left[ {%
\slashed{D}},G^{\dagger }\right] ,  \label{TF}
\end{equation}%
with $q$ the fermion charge -- one gets the most general form of the
electromagnetic potential:%
\begin{equation}
\slashed{A}=\left( 
\begin{array}{cc}
\gamma ^{\mu }A_{\mu }^{+} & \phi \gamma ^{5} \\ 
-\phi ^{\ast }\gamma ^{5} & \gamma ^{\mu }A_{\mu }^{-}%
\end{array}%
\right).  \label{pot}
\end{equation}

Thanks to seminal works on noncommutative geometry by Connes, followed by
other authors \cite{NC1,NC2,NC3,NC4,NC5,NC6,NC7,NC8}, attempts have been
made to derive the standard model of particle physics using a two-sheeted
space-time. In this context, the scalar field was associated with the Higgs
field. However, in the present study, one does not consider such a
hypothesis. Instead, one refers to the interpretation of the scalar field as
demonstrated in our previous works, where the $M_{4}\times Z_{2}$ approach
is derived as an effective limit of a two-brane world in a continuous bulk 
\cite{M4xZ2}. Then, one can assume the presence of an extra dimensional
component of the electromagnetic gauge field $U(1)$ in the bulk, and $\phi$
(see Eq. (\ref{pot})) represents this additional component dressed by
fluctuating fermionic fields in the bulk \cite{M4xZ2}. However, as a proof
of principle, in the present model one uses the definition of the field
strength used by Connes \textit{et al}. \cite%
{NC1,NC2,NC3,NC4,NC5,NC6,NC7,NC8}, one sets: 
\begin{equation}
\mathcal{F}=\left\{ i{\slashed{D}},\slashed{A}\right\} +e\slashed{A}%
\slashed{A},  \label{fstr}
\end{equation}
modulo the \textit{junk} terms \cite{NC1,NC2,NC3,NC4,NC5,NC6,NC7,NC8}, with $%
e$ here the electromagnetic coupling constant. The gauge field Lagrangian
being defined as: $\mathcal{L}=-\frac{1}{4}$Tr$\left\{ \mathcal{FF}\right\} $%
, from Eq. (\ref{fstr}) one gets \cite{NC1,NC2,NC3,NC4,NC5,NC6,NC7,NC8}:

\begin{eqnarray}
\mathcal{L} &=&-\frac{1}{4}F^{+\mu \nu }F_{\mu \nu }^{+}-\frac{1}{4}F^{-\mu
\nu }F_{\mu \nu }^{-}  \label{lagEM} \\
&&+\left( \mathcal{D}_{\mu }h\right) ^{\ast }\left( \mathcal{D}^{\mu
}h\right) -\frac{e^{2}}{2}\left( \left\vert h\right\vert ^{2}-2\eta
^{2}\right) ^{2},  \notag
\end{eqnarray}%
with $F_{\mu \nu }^{\pm }=\partial _{\mu }A_{\nu }^{\pm }-\partial _{\nu
}A_{\mu }^{\pm }$ ($A_{\mu }^{\pm }$ are the electromagnetic four-potentials
on each brane $(\pm ))$ and where the Lorenz gauge and the field
transversality are imposed, and where one has set:%
\begin{equation}
\mathcal{D}_{\mu }=\partial _{\mu }-ie\left( A_{\mu }^{+}-A_{\mu
}^{-}\right) ,  \label{covd}
\end{equation}%
and \cite{NC1,NC2,NC3,NC4,NC5,NC6,NC7,NC8}:%
\begin{equation}
h=\sqrt{2}\left( \phi +i\eta \right) ,  \label{hf}
\end{equation}%
with $\eta =g/e$. $h$ is a scalar field with a quartic self-interaction,
such that a vacuum state $h_{0}$ is characterized by: 
\begin{equation}
h_{0}=\eta \sqrt{2}e^{i\theta },  \label{val}
\end{equation}%
i.e. up to a phase $\theta $, the nature of which will be clarified in the
next section.

\section{Vaccum state phase and fluctuations}

\label{vacfluc}

Before proceeding, it is necessary to discuss the outcomes arising from the
dynamics of the field $h$ around a vacuum state $h_{0}$. The fluctuations of 
$h $ around $h_{0}$ can be conveniently described by introducing the
auxiliary fields $\left( \varphi ,\theta \right)$, such that: 
\begin{equation}
h=\sqrt{2}\left( \eta +\varphi /2\right) e^{i\theta }.  \label{A2}
\end{equation}%
Regarding the auxiliary fields $\left( \varphi ,\theta \right)$, the
electromagnetic gauge transformation (\ref{TF}) can be written as:\footnote{%
From the gauge transformation rule (\ref{TF}), the electromagnetic vector
potentials follow the usual transformation rule: $A_{\mu }^{\pm ^{\prime
}}=A_{\mu }^{\pm }+\partial _{\mu }\Lambda _{\pm }$, and the field $h$ follows the gauge transformation
rule: $h^{\prime }=h\exp (ie\left( \Lambda _{+}-\Lambda _{-}\right) )$. The
transformations (\ref{GT}) are equivalent to this gauge
transformation for the field $h$.}

\begin{equation}
\left\{ 
\begin{array}{c}
\varphi ^{\prime }=\varphi \\ 
\theta ^{\prime }=\theta +e\left( \Lambda _{+}-\Lambda _{-}\right)%
\end{array}%
\right. .  \label{GT}
\end{equation}%
Using now Eq. (\ref{A2}), the gauge covariant derivative (\ref{covd}) of $h$
in the Lagrangian (\ref{lagEM}) becomes: 
\begin{eqnarray}
\mathcal{D}_{\mu }h &=&\sqrt{2}e^{i\theta }\left( \frac{1}{2}\left( \partial
_{\mu }\varphi \right) \right.  \label{Dhp} \\
&&\left. +i\left( \eta +\varphi /2\right) \left( \left( \partial _{\mu
}\theta \right) -e\left( A_{\mu }^{+}-A_{\mu }^{-}\right) \right) \right) . 
\notag
\end{eqnarray}%
The Goldstone boson field $\theta $ could be eliminated by a
Brout-Englert-Higgs mechanism \cite{hi1,hi2,hi3} but then would lead to a
photon mass -- in the Lagrangian (\ref{lagEM}) -- that is difficult to
reconcile with current observations (see \cite{PDG} and references within).
Another possible mechanism -- i.e. gauge choice -- is a dynamical
compensation of the fluctuations of the field $\theta $ by the photon fields 
$A_{\mu }^{\pm }$ such that:%
\begin{equation}
\theta =e\int \left( A_{\mu }^{+}-A_{\mu }^{-}\right) dx^{\mu },  \label{ksi}
\end{equation}%
making $\theta $ an effective degree of freedom, driven by the photon fields 
$A_{\mu }^{\pm }$, with Eq. (\ref{ksi}) verifying the gauge transformations (%
\ref{GT}) and (\ref{TF}). Then the Lagrangian (\ref{lagEM}) becomes: 
\begin{eqnarray}
\mathcal{L} &=&-\frac{1}{4}F^{+\mu \nu }F_{\mu \nu }^{+}-\frac{1}{4}F^{-\mu
\nu }F_{\mu \nu }^{-}  \notag \\
&&+\frac{1}{2}\left( \partial _{\mu }\varphi \right) \left( \partial ^{\mu
}\varphi \right) -\frac{1}{2}m_{\varphi }^{2}\varphi ^{2},  \label{Laeff}
\end{eqnarray}
with\footnote{%
For the sake of clarity, we omitted the contributions $-(1/2)em_{\varphi
}\varphi ^{3}$ and $-(1/8)e^{2}\varphi ^{4}$ in Eq. (\ref{Laeff}) since we
consider the small fluctuations such that $\varphi \ll \eta $. These terms
could obviously be reintroduced as corrections.} $m_{\varphi }=2g.$ As a
result, the scalar field $\varphi$ describes a new massive scalar boson. 
In the following, the fluctuations $\varphi $ of
the field $h$ can be neglected as $h$ is dominated by $\eta $. At most, the
effective number of degrees of freedom will increase by one unit -- due to
the scalar boson -- without significant impact in the rest of our analysis.
In the following sections, without loss of generality and for illustrative
purpose, the phase $\theta$ will be considered as constant.

\section{Scalar field-induced C/CP violation and interbrane coupling
Hamiltonian}

\label{dys}

Writing now the two-brane Dirac equation including the gauge field from Eqs.
(\ref{LM4Z2}), (\ref{TF}) and (\ref{pot}) one gets:%
\begin{equation}
\left( 
\begin{array}{cc}
i\gamma ^{\mu }\left( \partial _{\mu }+iqA_{\mu }^{+}\right) -m & 
ig_{c}\gamma ^{5} \\ 
ig_{c}^{\ast }\gamma ^{5} & i\gamma ^{\mu }\left( \partial _{\mu }+iqA_{\mu
}^{-}\right) -m%
\end{array}%
\right) \Psi =0,  \label{DiracEM}
\end{equation}%
with:%
\begin{equation}
g_{c}=g+iq\phi _{0},  \label{gc}
\end{equation}%
here with $\phi _{0}=\eta \left( e^{i\theta }-i\right) $ (see Eqs. (\ref{hf}%
) and (\ref{val})) as the scalar field is on a vacuum state. Indeed, small
perturbations $\varphi $ $\left( \varphi \ll \eta \right) $ around the
vacuum state do not affect the baryogenesis model and correspond to a scalar
field propagating along the branes. It must be underlined that in our
previous work \cite{M4xZ2}, the role of the scalar field was neglected --
such that $g_{c}=g_{c}^{\ast }=g$ -- while here one explores its
consequences. It is then convenient to write $g_{c}$ as: 
\begin{equation}
g_{c}=\mathfrak{g}e^{i\alpha },  \label{gcp}
\end{equation}%
with:%
\begin{equation}
\mathfrak{g}=g\sqrt{1+2z\left( 1+z\right) \left( 1-\sin \theta \right) },
\label{gefff}
\end{equation}%
where $z=q/e$, and:%
\begin{equation}
\tan \alpha =\frac{z\cos \theta }{1+z\left( 1-\sin \theta \right) }.
\label{phase}
\end{equation}%
Then, thanks to a simple phase rescaling $\Psi \rightarrow T\Psi $, with $T=$
diag$\left\{ e^{i\alpha /2},e^{-i\alpha /2}\right\} $, one gets from Eq. (%
\ref{DiracEM}): 
\begin{equation}
\left( 
\begin{array}{cc}
i\gamma ^{\mu }\left( \partial _{\mu }+iqA_{\mu }^{+}\right) -m & i\mathfrak{%
g}\gamma ^{5} \\ 
i\mathfrak{g}\gamma ^{5} & i\gamma ^{\mu }\left( \partial _{\mu }+iqA_{\mu
}^{-}\right) -m%
\end{array}%
\right) \Psi =0.  \label{DiracR}
\end{equation}%
Then, $\mathfrak{g}$ becomes the effective coupling constant between the
visible and the hidden sectors for the fermion dressed by the scalar field.\
Now, let us consider the standard procedure for obtaining the Pauli equation
from the Dirac equation in its two-brane formulation (\ref{DiracR}). By
doing so, one can derive the interbrane coupling Hamiltonian for a fermion
(see \cite{M4xZ2, pheno}):

\begin{equation}
\mathcal{W}=\varepsilon \left( 
\begin{array}{cc}
0 & \mathbf{u} \\ 
\mathbf{u}^{\dag } & 0%
\end{array}%
\right) ,  \label{hcm}
\end{equation}%
where:%
\begin{equation}
\varepsilon =\mathfrak{g}\mu \left\vert \mathbf{A}_{+}-\mathbf{A}%
_{-}\right\vert ,  \label{eps}
\end{equation}
with $\mathbf{A}_{\pm }$ the local magnetic vector potentials in each brane 
\cite{M4xZ2,pheno}, $\mu $ the magnetic moment of the fermion and $\mathbf{u}
$ a unitary matrix such that: $\mathbf{u=}i\mathbf{e}\cdot \mathbf{\sigma }$
with $\mathbf{e}=\left( \mathbf{A}_{+}-\mathbf{A}_{-}\right) /\left\vert 
\mathbf{A}_{+}-\mathbf{A}_{-}\right\vert $. The phenomenology related to $%
\mathcal{W}$ is explored and is detailed elsewhere \cite%
{pheno,Exp0,Exp1,Exp2,Exp3,Exp4,PLB,M4xZ2,cp1,cp2}. From the Hamiltonian (%
\ref{hcm}), one can show that a particle should oscillate between two
states: One localized in our brane and the other localized in the hidden
world \cite{M4xZ2}. While such oscillations are suppressed for charged
particles \cite{pheno, char0, Dvali}, they remain possible for composite
particles with neutral charge such as neutrons or antineutrons \cite{pheno,
char0, Dvali}, for which the above coupling has the same form. This could
result in the disappearance \cite{Exp0} or reappearance of neutrons,
allowing for passing-through-walls neutron experiments, which have been
conducted in the last decade \cite{Exp1,Exp2,Exp3,Exp4}. Such phenomena
would appear as a baryon number violation.

The interbrane coupling Hamiltonian $\mathcal{W}$ for the anti-fermion can
be obtained through the charge conjugation $q\rightarrow -q$ in Eqs. (\ref%
{hcm}) and (\ref{gefff}). One labels $\overline{\mathfrak{g}}$ the coupling
constant between the visible and the hidden sectors for the anti-fermion.
For the antiparticle the sign change $\mu \rightarrow -\mu $ due to the
charge conjugation can be effectively eliminated through a relevant phase
rescaling in Eq. (\ref{hcm}). It is not the case for the coupling constant.
When $\phi =0$, we have $\mathfrak{g}=g$, and the antiparticle also exhibits 
$\overline{\mathfrak{g}}=g$. However, in the case where $\phi \neq 0$, one
finds $\mathfrak{g}\rightarrow \overline{\mathfrak{g}}\neq \mathfrak{g}$
(with $\overline{\mathfrak{g}},\mathfrak{g}>0$), and this disparity cannot
be canceled: the interbrane coupling magnitude differs between the particle
and the antiparticle. Then, the presence of a scalar field in the two-brane
universe breaks the symmetry between $\overline{\mathfrak{g}}$ and $%
\mathfrak{g}$. It must be underlined that such an asymmetry would be hidden
from us in our visible world, except for experiments involving neutron and
antineutron disappearance and/or reappearance \cite{Exp1,Exp2,Exp3,Exp4}.
The state of the art of this kind of experiment \cite%
{Exp0,Exp1,Exp2,Exp3,Exp4} for the neutron requires nuclear reactors, thus
implying there is few hope for convincing experiments using antineutrons.
Nevertheless, in section \ref{results}, one will suggest a way to get
observational constraints for the present scenario by testing other
consequences induced by the scalar field.

\section{Neutron and antineutron interbrane coupling constants}

\label{inter}

\begin{figure}[ht!]
\centerline{\ \includegraphics[width=8.5cm]{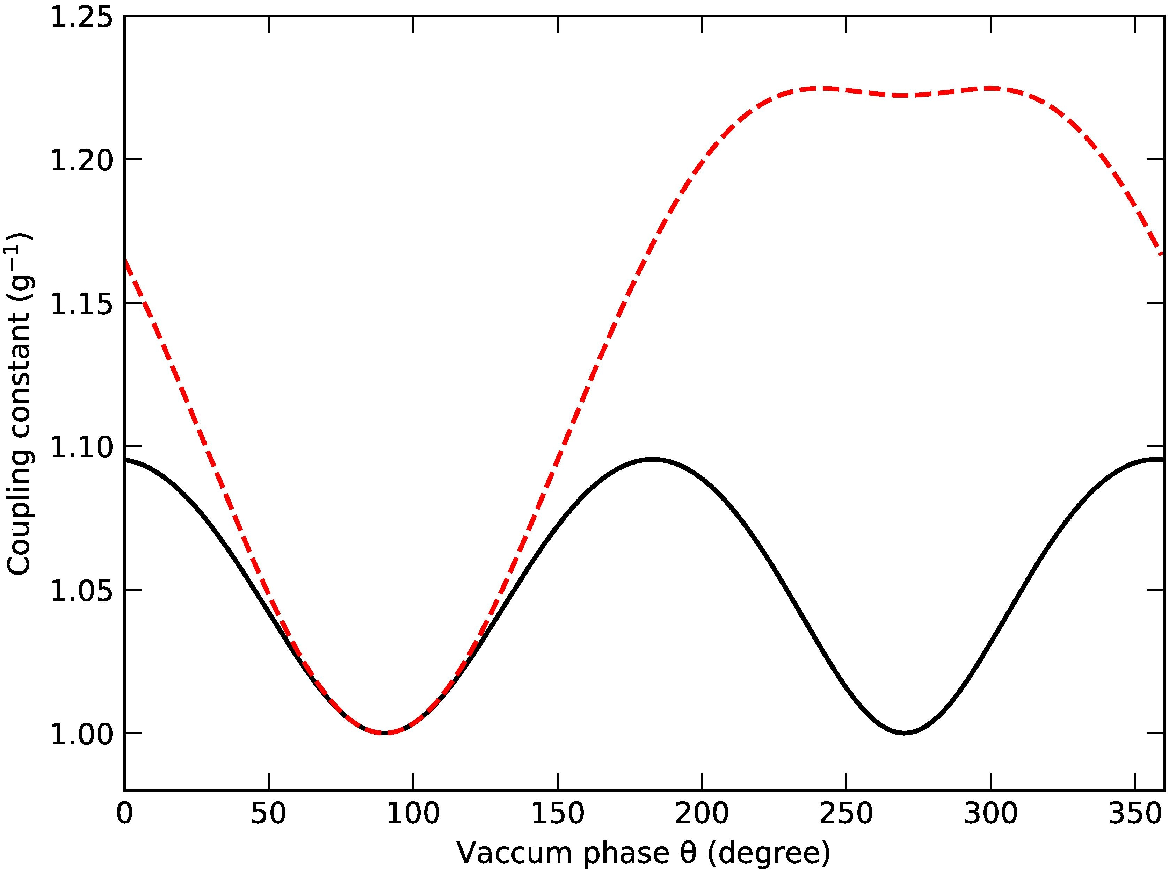}}
\caption{(Color online). Normalized coupling constant for neutron $\mathfrak{%
g}/g$ (black line) and antineutron $\overline{\mathfrak{g}}/g$ (red dashed
line) against the scalar field phase $\protect\theta $ in the vacuum state.}
\label{fig:f1}
\end{figure}

\begin{figure}[ht!]
\centerline{\ \includegraphics[width=8.5cm]{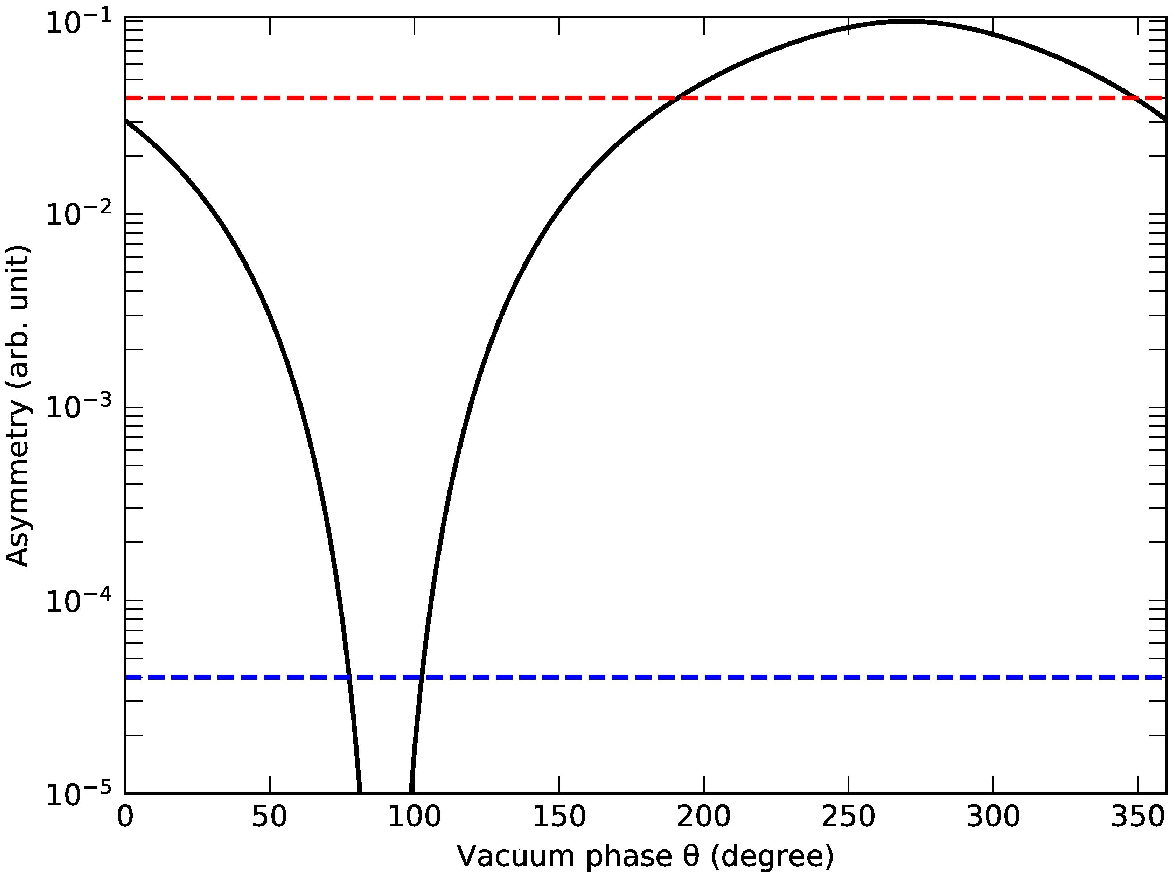}}
\caption{(Color online). Asymmetry $\protect\delta = \Delta \mathfrak{g/g}$
against the scalar field phase $\protect\theta $ in the vacuum state. Upper
red dashed line: upper limit on the asymmetry compatible with baryogenesis
as shown in section \protect\ref{results} (see Eq. (\protect\ref{ct})).
Lower blue dashed line: lower limit compatible with baryogenesis (section 
\protect\ref{results}, Eq. (\protect\ref{ct})).}
\label{fig:f2}
\end{figure}

The two-brane Dirac equation (\ref{DiracR}) can be fundamentally derived 
\cite{cp1,cp2} to describe quarks within baryons (or mesons). But, Eq. (\ref%
{gefff}) cannot be directly applied to characterize the neutron \cite%
{cp1,cp2} or the antineutron as they are not point-like particles. In order
to address this issue, the well-known quark constituent model \cite%
{CQMbook,CQM1,CQM2,CQM3} is pursued as outlined elsewhere \cite{cp1,cp2}. In
this context, assuming that $\mathfrak{g}$ (respectively $\widehat{\mathbf{%
\mu }}_{n}$) represents the coupling constant (respectively, the magnetic
moment operator) of the neutron, the quark constituent model \cite%
{CQMbook,CQM1,CQM2,CQM3} is employed and one gets: 
\begin{equation}
\mathfrak{g}\widehat{\mathbf{\mu }}_{n}=\sum\limits_{q}\widehat{\mathbf{\mu }%
}_{q}\mathfrak{g}_{q},  \label{guop}
\end{equation}%
where $\mathfrak{g}_{q}$ (respectively $\widehat{\mathbf{\mu }}_{q}$) refers
to the coupling constant (respectively the magnetic moment operator) of each
quark constituting the neutron with $\widehat{\mathbf{\mu }}%
_{n}=\sum\limits_{q}\widehat{\mathbf{\mu }}_{q}$. The magnetic moment of the
neutron $\mu _{n}$ is then calculated by taking the expectation value of the
operator $\widehat{\mathbf{\mu }}_{n}$, and one gets \cite{CQMbook}:%
\begin{equation}
\mu _{n}=\left\langle n,\uparrow \right\vert \widehat{\mathbf{\mu }}%
\left\vert n,\uparrow \right\rangle =\frac{4}{3}\mu _{d}-\frac{1}{3}\mu _{u},
\label{un}
\end{equation}%
where -- without loss of generality -- one has considered the neutron with
spin up such that \cite{CQMbook}:%
\begin{eqnarray}
\left\vert n,\uparrow \right\rangle  &=&\frac{1}{\sqrt{18}}\left(
-2\left\vert d,\uparrow \right\rangle \left\vert d,\uparrow \right\rangle
\left\vert u,\downarrow \right\rangle +\right.   \label{wf} \\
&&\left\vert d,\uparrow \right\rangle \left\vert d,\downarrow \right\rangle
\left\vert u,\uparrow \right\rangle +\left\vert d,\downarrow \right\rangle
\left\vert d,\uparrow \right\rangle \left\vert u,\uparrow \right\rangle  
\notag \\
&&\left. +\,\text{permutations}\right) ,  \notag
\end{eqnarray}%
with $\left\vert u,\updownarrow \right\rangle $ and $\left\vert
d,\updownarrow \right\rangle $ the quark up and the quark down
wave-functions respectively, either with spin up $\uparrow $ or down $%
\downarrow $. Also, one gets:%
\begin{equation}
\mu _{u}=\frac{2}{3}\frac{e\hbar }{2m_{u}}\text{ and }\mu _{d}=-\frac{1}{3}%
\frac{e\hbar }{2m_{d}}.  \label{uq}
\end{equation}%
Using $m_{u}=m_{d}=m_{Q}=327$ MeV \cite{CQMbook,CQM1,CQM2,CQM3}, one
obtains \cite{CQMbook}: 
\begin{equation}
\mu _{n}=-\frac{2}{3}\frac{e\hbar }{2m_{Q}}.  \label{une}
\end{equation}%
Doing the same for $\mathfrak{g}\widehat{\mathbf{\mu }}_{n}$, one deduces from Eq. (\ref%
{guop}): 
\begin{equation}
\mathfrak{g}\mu _{n}=\frac{4}{3}\mathfrak{g}_{d}\mu _{d}-\frac{1}{3}%
\mathfrak{g}_{u}\mu _{u},  \label{gu}
\end{equation}%
Next, one divides Eq. (\ref{gu}) by Eq. (\ref{une}), and one gets:%
\begin{equation}
\mathfrak{g}=\frac{2}{3}\mathfrak{g}_{d}+\frac{1}{3}\mathfrak{g}_{u}.
\label{geff}
\end{equation}%
From Eqs. (\ref{gefff}) and (\ref{geff}), one deduces the explicit
expression for the coupling constant $\mathfrak{g}$ between the visible and
the hidden sectors:%
\begin{eqnarray}
\frac{\mathfrak{g}}{g} &=&\frac{2}{9}\sqrt{5+4\sin \theta }  \label{geffsg}
\\
&&+\frac{1}{9}\sqrt{29-20\sin \theta }.  \notag
\end{eqnarray}%
Doing the same for the antineutron, one gets the related coupling constant $%
\overline{\mathfrak{g}}$ between the visible and the hidden sectors:%
\begin{eqnarray}
\frac{\overline{\mathfrak{g}}}{g} &=&\frac{2}{9}\sqrt{17-8\sin \theta }
\label{geffsgAP} \\
&&+\frac{1}{9}\sqrt{5+4\sin \theta }.  \notag
\end{eqnarray}%
In the following, one defines the asymmetry of the interbrane coupling
constants of the neutron and antineutron as: 
\begin{equation}
\delta =\frac{\Delta \mathfrak{g}}{\mathfrak{g}}=\frac{\left\vert \overline{%
\mathfrak{g}}-\mathfrak{g}\right\vert }{\overline{\mathfrak{g}}+\mathfrak{g}}%
,  \label{asydef}
\end{equation}%
and one gets: 
\begin{equation}
\delta =\frac{\left\vert \sqrt{5+4\sin \theta }+\sqrt{29-20\sin \theta }-2%
\sqrt{17-8\sin \theta }\right\vert }{3\sqrt{5+4\sin \theta }+\sqrt{29-20\sin
\theta }+2\sqrt{17-8\sin \theta }},  \label{var}
\end{equation}%
which does not depend on the expression of $g$ and therefore, not on the
bulk dimensionality. In Fig. \ref{fig:f1}, the normalized coupling constants
for the neutron, $\mathfrak{g}/g$, and the antineutron, $\overline{\mathfrak{%
g}}/g$, are illustrated against the scalar field phase $\theta $ in the
vacuum state from Eqs. (\ref{geffsg}) and (\ref{geffsgAP}). In a same way,
Fig. \ref{fig:f2} displays the asymmetry $\Delta \mathfrak{g/g}$ plotted
against $\theta $ from Eq. (\ref{var}). The upper red and lower blue dashed
lines bound the values of the asymmetry $\delta $, which are compatible with
the observed imbalance of the baryon-antibaryon populations today. This will
be shown and discussed in section \ref{results} (see Eq. (\ref{ct})).

\section{Baryon phenomenology in the early two-brane Universe}

\label{model}

Usually, the Boltzmann transport equation \cite{Bolt2, PU} leads to the
Lee-Weinberg equations \cite{LW} that govern the density of relic particles
in the expanding Universe. The density of baryons $n_{B}$ (respectively
antibaryons $n_{\overline{B}}$) thus obeys to \cite{Bolt2, PU}:%
\begin{equation}
\partial _{t}n_{B}+3Hn_{B}=-\left\langle \sigma _{a}v\right\rangle \left(
n_{B}n_{\overline{B}}-n_{B,eq}n_{\overline{B},eq}\right) ,  \label{BTE}
\end{equation}%
with $H$ the Hubble parameter, $\sigma _{a}$ the baryon-antibaryon
annihilation cross-section, $v$ the relative velocity between particles, and 
$\left\langle \cdots \right\rangle $ the thermal average at temperature $T$.
Quantities $n_{B,eq}$ and $n_{\overline{B},eq}$ are at the thermal
equilibrium and are described by the Fermi-Dirac statistics. Without
baryon-antibaryon asymmetry, one would have $n_{B}=n_{\overline{B}}$, and
the same expression would occur for antibaryons through the $%
n_{B}\leftrightarrow n_{\overline{B}}$ substitution. Under such conditions,
particles would simply annihilate until the expansion of space froze the
process by reducing the probability of collision between particles and
antiparticles. Then, baryons and antibaryons would have the same density in
the Universe (there would be no asymmetry) but lower by many orders of
magnitude than the current observed values. However, the current imbalance
in the observed Universe between baryons and antibaryons -- with a large
photon population -- suggests an early asymmetry.\ One actually observes 
\cite{Bolt}:%
\begin{equation}
Y_{B}-Y_{\overline{B}}=\left( 8.8\pm 0.6\right) \times 10^{-11},
\label{assy}
\end{equation}%
where $Y_{X}=n_{X}/s$ is the comoving particle density, i.e. the particle
density $n_{X}$ related to the entropy density $s$, itself proportional to
the photon population \cite{Bolt2, PU}. As the temperature of the Universe
decreased, a baryonic asymmetry could have precluded the complete
annihilation of all matter and antimatter, resulting in a very small excess
of matter over antimatter. The baryogenesis process supposes that the three
Sakharov conditions \cite{Cond} are satisfied: Baryon number violation,
C-symmetry and CP-symmetry violation, and interactions out of thermal
equilibrium. Currently, C/CP violation processes known in physics are too
weak in magnitude to explain baryogenesis, and solutions are expected from
attempts to build a grand unified theory. However, for now, the origin of
the imbalance between matter and antimatter is still unknown, despite the
existence of many hypotheses \cite{baryoR,Bolt,rev1,BaryoG}.

In previous sections, it was underlined that
neutron and antineutron could be the portal inducing the baryogenesis right
after the phase transition from quark-gluon plasma to hadron gas (QGPHG).
Keeping the Sakharov conditions in mind, we propose to discuss the magnitude
of the asymmetry between $\mathfrak{g}$ and $\overline{\mathfrak{g}}$ and
its consequences in a baryogenesis scenario. Between the QGPHG transition ($%
T_{0}\approx 160$ MeV) and the end of baryon-antibaryon annihilation ($%
T\approx 20$ MeV), we need to explain the similarities of the temperatures
in each brane, a condition necessary as shown later. This could be possible
if the branes had collided during the initial stage of the Big Bang,
regardless of the underlying mechanisms during the collision of the branes 
\cite{Branerev,cob1,cob2,cob3,cob4,cob5,cob6,cob7,cob8,cob9,cob10,cob11}.

Let us consider matter (or antimatter) exchange between two branes: the one
corresponding to our visible Universe and a hidden one. The process is
described through the Hamiltonian (\ref{hcm}) added to a Hamiltonian $%
\mathcal{H}_{0}$ describing the neutron (or antineutron) in each brane such
that:%
\begin{equation}
\mathcal{H}=\mathcal{H}_{0}+\mathcal{W},  \label{hami}
\end{equation}
with $\mathcal{H}_{0}=$ diag$\left\{ E_{+},E_{-}\right\} $ and $E_{\pm
}=E_{0,\pm }+V_{F,\pm }$, where $E_{0,\pm \text{ }}$are the eigenenergies of
the particle in vacuum either in its visible state or its hidden state due
to the gravitational potentials of each brane, and $V_{F,\pm }$ are the
Fermi potentials of the materials through which the particle travels \cite%
{Exp1,Exp4}. The visible or hidden states of matter (or antimatter) are
quantum states, but not eigenstates of (\ref{hcm}). Therefore, the Lindblad
equation formalism \cite{Lindblad} is necessary to describe the dynamics of
quantum states that change a visible neutron $n$ into a hidden one $%
n^{\prime }$ (or a visible antineutron $\overline{n}$ into a hidden one $%
\overline{n}^{\prime }$) -- and vice versa -- as a result of interactions
with many scatterers $X$ (i.e. $n+X\leftrightarrow n^{\prime }+X$). This
equation extends the Liouville-Von Neumann equation related to the density
matrix $\rho $ -- and allows the study of the evolution of a quantum system
(the neutron or antineutron) interacting with two environments that are not
in thermal equilibrium \cite{Lindblad}, i.e., a set of scatterers $X$ in our
brane and a set of scatterers $X^{\prime }$ in the hidden brane. For the
two-brane Universe, the Lindblad equation can be written as:%
\begin{equation}
\partial _{t}\rho +\frac{3}{2}\left\{ H,\rho \right\} =i\left[ \rho ,%
\mathcal{H}\right] +L(\rho ),  \label{Lindblad}
\end{equation}%
where $\left\{ A,B\right\} =AB+BA$ defines the anticommutator,\footnote{%
The term $(3/2)\left\{ H,\rho \right\} $ arises from the covariant
derivatives in the Dirac equation for a universe with two space-time sheets
(or branes) endowed with their own tensor metric: $g_{\pm ,\mu \nu
}^{(4)}=\, $diag$(1,-a_{\pm }^{2}(t),-a_{\pm }^{2}(t),-a_{\pm }^{2}(t))$
with scale factors $a_{\pm }$ such that $H_{\pm }=\left( \partial _{t}a_{\pm
}\right) /a_{\pm }$ are the Hubble parameters in each brane.} with $H=$ diag$%
\left\{ H_{+},H_{-}\right\} $ and $H_{\pm }$ the Hubble parameters in each
brane. The Lindblad operator $L(\rho )$ is defined as \cite{Lindblad}:%
\begin{equation}
L(\rho )=\sum_{m}\Gamma _{m}\left( C_{m}\rho C_{m}^{\dagger }-\frac{1}{2}%
\left\{ \rho ,C_{m}^{\dagger }C_{m}\right\} \right) ,  \label{L}
\end{equation}%
where $C_{m}$ ($m=\pm $) are the jump operators describing the wave function
reduction process either into the visible or into the hidden branes when the
system interacts with its environment.\footnote{$C_{+}=$ diag$\left\{
1,0\right\} $ and $C_{-}=$ diag$\left\{ 0,1\right\} .$} Then, $\Gamma _{+}$
(respectively $\Gamma _{-}$) describes the collisional rate between the
neutron (or antineutron) and the environment in the brane $+$ (respectively
in the brane $-$) when it is assumed to be in this brane. In the following, $%
T$ is the temperature in our visible braneworld and $T^{\prime }$ in the
hidden braneworld, such that:

\begin{equation}
\kappa =\frac{T}{T^{\prime }},  \label{kappa}
\end{equation}%
where $\kappa $ is a constant parameter. Setting $\sigma $ the usual elastic
cross-section $\sigma =\sigma (n+X\longrightarrow n+X)$, one gets: $\Gamma
_{+}$ $=\left\langle \sigma v\right\rangle n_{X}$ and $\Gamma _{-}$ $%
=\left\langle \sigma v\right\rangle ^{\prime }n_{X^{\prime }}$,\footnote{$%
\left\langle \cdots \right\rangle ^{\prime }$ is the thermal average at $%
T^{\prime }.$} where \cite{sg}: 
\begin{eqnarray}
\left\langle \sigma v\right\rangle &=&\int \int d^{3}\mathbf{v}_{1}d^{3}%
\mathbf{v}_{2}f_{T}\left( \mathbf{v}_{1}\right) f_{T}\left( \mathbf{v}%
_{2}\right) \sigma \left\vert \mathbf{v}_{1}-\mathbf{v}_{2}\right\vert 
\notag \\
&=&\frac{x^{3/2}}{2\sqrt{\pi }}\int_{0}^{\infty }dvv^{2}e^{-xv^{2}/4}\sigma
v,  \label{tav}
\end{eqnarray}%
with $x=m/T$ the usual parameter \cite{Bolt2, PU} used to follow the
primordial particle dynamics, and $m$ a mass reference, here equals to the
typical mass of the nucleon: $939$ MeV/c$^{2}$. One also uses $x^{\prime
}=m/T^{\prime }=\kappa x$.

Setting:%
\begin{equation}
\rho =\left( 
\begin{array}{cc}
\rho _{+} & x-iy \\ 
x+iy & \rho _{-}%
\end{array}%
\right) ,  \label{rho}
\end{equation}%
Eq. (\ref{Lindblad}) for unpolarized fermions becomes:%
\begin{equation}
\left\{ 
\begin{array}{c}
\partial _{t}\rho _{+}=-3H_{+}\rho _{+}+2\varepsilon y \\ 
\partial _{t}\rho _{-}=-3H_{-}\rho _{-}-2\varepsilon y \\ 
\partial _{t}x=-\left( 3H+\Gamma \right) x-\Delta Ey \\ 
\partial _{t}y=-\left( 3H+\Gamma \right) y+\Delta Ex-\varepsilon \left( \rho
_{+}-\rho _{-}\right)%
\end{array}%
\right.  \label{sys}
\end{equation}%
with $\Delta E=E_{+}-E_{-}$, \ $H=(H_{+}+H_{-})/2$ and $\Gamma =\left(
\Gamma _{+}+\Gamma _{-}\right) /2$ and where $\Delta E$, $\Gamma $ and $%
\varepsilon $ can depend on time. Of course, $\varepsilon $ is given by Eq. (%
\ref{eps}) where the coupling constant $\mathfrak{g}$ between the visible
and the hidden sectors of the neutron acts (see sections \ref{dys} and \ref%
{inter}). Here, due to the isotropy and the homogeneity of the Universe in
both branes, and due to the strong collisional dynamics:\footnote{%
The Fermi potential writes as $V_{F}=(2\pi \hbar ^{2}/m)bn_{X}$ with $m$ the
neutron mass and $b$ the scattering length on a free nucleon ($b\approx 0.73$
fm). Then, $\Gamma \gg V_{F}$ leads to $\left\langle \sigma v\right\rangle
\gg (2\pi \hbar /m)b$ which is verified in the present work.} $\Gamma \gg
H>\Delta E$. This allows for the stationary phase approximation \cite%
{Exp1,Exp3,Exp4}: $\partial _{t}x\approx \partial _{t}y\approx 0$, and the
system (\ref{sys}) can be conveniently recast as:%
\begin{equation}
\left\{ 
\begin{array}{c}
\partial _{t}n_{n}+3H_{+}n_{n}=-\gamma \left( n_{n}-n_{n^{\prime }}\right)
\\ 
\partial _{t}n_{n^{\prime }}+3H_{-}n_{n^{\prime }}=-\gamma \left(
n_{n^{\prime }}-n_{n}\right)%
\end{array}%
\right. ,  \label{sys2}
\end{equation}%
with $\gamma $ the neutron transition rate between branes such that:%
\begin{equation}
\gamma =\frac{2\left( 3H+\Gamma \right) \varepsilon ^{2}}{\left( 3H+\Gamma
\right) ^{2}+\Delta E^{2}}.  \label{gamma}
\end{equation}%
and where one used: $n_{n}=n_{0}\rho _{+}$ and $n_{n^{\prime }}=n_{0}\rho
_{-}$ with $n_{0}$ the global neutron population in the two-brane Universe 
\cite{Exp1,Exp4}. Since $\Gamma \gg H>\Delta E$, one gets: $\gamma \sim
2\varepsilon ^{2}/\Gamma .$

During the period of interest, the coupling parameter $\varepsilon $ depends
only on the typical amplitude $A$ of the magnetic vector potentials related
to primordial magnetic fields \cite{magf}, then:\footnote{%
The magnetic vector potential is given by $A_{0}\sim B_{0}L_{0}$ with $%
B_{0}\approx 10^{4}$ T the field strength at the QCD phase transition time
(i.e. at $T_{0}$) \cite{magf2} and $L_{0}$ the maximal coherence length of
the magnetic field at the same epoch, i.e. $L_{0}\sim H^{-1}$ \cite{magf2}
with $H$ the Hubble parameter.} $A=A_{0}(x_{0}/x)$, with $A_{0}\approx
4.0\times 10^{8}$ T.m the typical amplitude at $T=T_{0}$, i.e. at the QGPHG
transition \cite{magf,magf2}. Then: 
\begin{equation}
\varepsilon =\varepsilon _{0}\frac{x_{0}}{x},  \label{ccd}
\end{equation}%
with\footnote{%
Since $\varepsilon =\mathfrak{g} \mu _{n}\left\vert \mathbf{A}_{+}-\mathbf{A}%
_{-}\right\vert $, one considers that: $A_{+}=A_{0}(x_{0}/x)$ and $%
A_{-}=A_{0}(x_{0}/x^{\prime })$ and the fact that $\mathbf{A}_{+}$ and $%
\mathbf{A}_{-}$ should have different orientations in various domains of the
early Universe. Then, one uses $\varepsilon =\mathfrak{g} \mu
_{n}\left\langle \left\vert \mathbf{A}_{+}-\mathbf{A}_{-}\right\vert
\right\rangle $ with $\left\langle \left\vert \mathbf{A}_{+}-\mathbf{A}%
_{-}\right\vert \right\rangle $ the averaged value over all the possible
relative directions between $\mathbf{A}_{+}$ and $\mathbf{A}_{-}$. One
shows: $\left\langle \left\vert \mathbf{A}_{+}-\mathbf{A}_{-}\right\vert
\right\rangle =A_{+}(2/\pi )\left( 1+1/\kappa \right) E\left( \frac{4\kappa 
}{\left( 1+\kappa \right) ^{2}}\right) \sim A_{+}$ for $1<\kappa <3$. $E(x)$
is the complete elliptic integral of the second kind.} $\varepsilon _{0}=%
\mathfrak{g} \mu _{n}A_{0}$.

For antineutrons, a set of equations similar to Eq. (\ref{sys2}) can be
derived -- with $n_{\overline{n}}$ and $n_{\overline{n}^{\prime }}$ -- but
where $\overline{\gamma }=2\overline{\varepsilon }^{2}/\overline{\Gamma }$
-- with $\overline{\varepsilon }_{0}=\overline{\mathfrak{g}}\mu _{\overline{n%
}}A_{0}$ and where $\overline{\Gamma }$ will be conveniently defined in
details below. $\overline{\mathfrak{g}}$ is of course the coupling constant
between the visible and the hidden sectors for the anti-neutron as defined
in sections \ref{dys} and \ref{inter}.

The system of equations (\ref{sys2}) now allows us to extend Eq. (\ref{BTE}%
). The right-hand side of equation (\ref{BTE}) for neutrons (or
antineutrons) can be written for both brane $+$ and brane $-$ and must be
added to the right-hand sides of the two expressions in system (\ref{sys2})
for each brane.

In the period of interest, the Universe is composed of various baryons,
mesons, leptons, and neutrinos. However, we consider that the dynamics of
nucleons primarily depends on their equilibrium with the lightest leptons
and related neutrinos. Electrons, positrons, neutrinos, and antineutrinos
are relativistic and in thermal equilibrium with the photon bath. Therefore: 
$n_{e^{-}}=n_{e^{-},eq}=n_{e^{+}}=n_{e^{+},eq}=n_{l,eq}$ (the same is true
for the hidden brane). At equilibrium, above the threshold temperature of
the electron-positron plasma, the populations of protons and neutrons
follow: $n_{n,eq}=n_{p,eq}\left( m_{n}/m_{p}\right) ^{3/2}\exp (-\Delta m/T)$
(with $\Delta m=m_{n}-m_{p}$) as neutrons contribute to the protons
population mainly through $n+e^{+}\rightarrow p+\overline{\nu }$ and as
protons contributes to the neutrons population through $p+e^{-}\rightarrow
n+\nu $. During the period of interest, as a fair approximation, we assume: $%
n_{p,eq}=n_{n,eq}$ and $n_{\overline{p},eq}=n_{\overline{n},eq}$ and the
same for the hidden brane, but also $n_{n}=n_{p}=(1/2)n_{B}$, $n_{\overline{n%
}}=n_{\overline{p}}=(1/2)n_{\overline{B}}$, $n_{n^{\prime }}=n_{p^{\prime
}}=(1/2)n_{B^{\prime }}$ and $n_{\overline{n}^{\prime }}=n_{\overline{p}%
^{\prime }}=(1/2)n_{\overline{B}^{\prime }}$. Writing then the system (\ref%
{sys2}) including the Lee-Weinberg equations for each particle species --
and for particles and antiparticles -- and assuming the above hypothesis,
one easily obtains:%
\begin{eqnarray}
\frac{dY_{B}}{dx} &=&-\left\langle \sigma _{B\overline{B},a}v\right\rangle
\eta \frac{s}{Hx}\left( Y_{B}Y_{\overline{B}}-Y_{B,eq}Y_{\overline{B}%
,eq}\right)  \notag \\
&&-(1/2)\frac{\gamma \eta }{Hx}\left( Y_{B}-Y_{B^{\prime }}\right) ,
\label{B1}
\end{eqnarray}%
\begin{eqnarray}
\frac{dY_{\overline{B}}}{dx} &=&-\left\langle \sigma _{B\overline{B}%
,a}v\right\rangle \eta \frac{s}{Hx}\left( Y_{B}Y_{\overline{B}}-Y_{B,eq}Y_{%
\overline{B},eq}\right)  \notag \\
&&-(1/2)\frac{\overline{\gamma }\eta }{Hx}\left( Y_{\overline{B}}-Y_{%
\overline{B}^{\prime }}\right) ,  \label{B2}
\end{eqnarray}%
\begin{eqnarray}
\frac{dY_{B^{\prime }}}{dx} &=&-\left\langle \sigma _{B\overline{B}%
,a}v\right\rangle ^{\prime }\eta ^{\prime }\frac{\kappa s^{\prime }}{%
H^{\prime }x^{\prime }}\left( Y_{B^{\prime }}Y_{\overline{B}^{\prime
}}-Y_{B^{\prime },eq}Y_{\overline{B}^{\prime },eq}\right)  \notag \\
&&-(1/2)\frac{\gamma \kappa \eta ^{\prime }}{H^{\prime }x^{\prime }}\left(
Y_{B^{\prime }}-Y_{B}\right) ,  \label{B3}
\end{eqnarray}%
\begin{eqnarray}
\frac{dY_{\overline{B}^{\prime }}}{dx} &=&-\left\langle \sigma _{B\overline{B%
},a}v\right\rangle ^{\prime }\eta ^{\prime }\frac{\kappa s^{\prime }}{%
H^{\prime }x^{\prime }}\left( Y_{B^{\prime }}Y_{\overline{B}^{\prime
}}-Y_{B^{\prime },eq}Y_{\overline{B}^{\prime },eq}\right)  \notag \\
&&-(1/2)\frac{\overline{\gamma }\kappa \eta ^{\prime }}{H^{\prime }x^{\prime
}}\left( Y_{\overline{B}^{\prime }}-Y_{\overline{B}}\right) ,  \label{B4}
\end{eqnarray}%
where we have introduced the comoving particle densities: $Y_{B}=n_{B}/s$, $%
Y_{\overline{B}}=n_{\overline{B}}/s$, $Y_{B^{\prime }}=n_{B^{\prime
}}/s^{\prime }$ and $Y_{\overline{B}^{\prime }}=n_{\overline{B}^{\prime
}}/s^{\prime }$ with $s$ and $s^{\prime }$ the entropy densities in each
brane. We have also proceeded to the variable changing $t\rightarrow x$ such
that $\left( H_{+},H_{-}\right) \rightarrow \left( H,H^{\prime }\right) $
(see Eq. \ref{Hubble}) with the relations \cite{Bolt2, PU}: $dx/dt=Hx/\eta $
and $dx^{\prime }/dt=H^{\prime }x^{\prime }/\eta ^{\prime }$ in each brane,
where:%
\begin{equation}
\eta =1-\frac{x}{3q_{\ast }}\frac{dq_{\ast }}{dx},  \label{ad}
\end{equation}%
with $q_{\ast }$ the effective number of degrees of freedom defined for the
entropy density such that \cite{Bolt2, PU}: 
\begin{equation}
s=\frac{2\pi ^{2}}{45}m^{3}q_{\ast }x^{-3}.  \label{s}
\end{equation}%
While $\eta $ is often close to $1$ during most of the radiation era, it is
not the case shortly after the QGPHG transition as pions and muons
annihilate between $160$ MeV and $100$ MeV leading then to a fast change of $%
q_{\ast }$ against $x$. In the same way, since the period of interest is
radiatively-dominated, the Hubble parameter is defined through \cite{Bolt2,
PU}:%
\begin{equation}
H=\frac{2\pi \sqrt{\pi }}{3\sqrt{5}}\frac{m^{2}}{M_{P}}g_{\ast }^{1/2}x^{-2},
\label{Hubble}
\end{equation}%
with $g_{\ast }$ the effective number of degrees of freedom defined for the
energy density, and where $M_{P}$ is the Planck mass. Both functions $%
g_{\ast }$ and $q_{\ast }$ can be fitted from exact computations \cite{DF}
and one can set $g_{\ast }=q_{\ast }$ \cite{Bolt2, PU,DF}. The equilibrium
state of the comoving particle densities is defined as \cite{Bolt2, PU}: 
\begin{equation}
Y_{X,eq}=\frac{45}{2\pi ^{4}}\sqrt{\frac{\pi }{8}}\frac{g_{X}}{q_{\ast }}%
x^{3/2}e^{-x},  \label{eq}
\end{equation}%
In the above equations (\ref{B1}) to (\ref{B4}), $\left\langle \sigma _{B%
\overline{B},a}v\right\rangle $ and $\left\langle \sigma _{B\overline{B}%
,a}v\right\rangle ^{\prime }$ appear as the average rate of
baryon-antibaryon annihilation with: $\sigma _{B\overline{B},a}=(1/4)\left(
\sigma _{n\overline{n},a}+\sigma _{p\overline{p},a}+\sigma _{n\overline{p}%
,a}+\sigma _{p\overline{n},a}\right) $. One also defines: 
\begin{eqnarray}
2\Gamma &=&\left\langle \sigma _{BB}v\right\rangle sY_{B}+\left\langle
\sigma _{B\overline{B}}v\right\rangle sY_{\overline{B}}  \label{G} \\
&&+\left\langle \sigma _{BB}v\right\rangle ^{\prime }s^{\prime }Y_{B^{\prime
}}+\left\langle \sigma _{B\overline{B}}v\right\rangle ^{\prime }s^{\prime
}Y_{\overline{B}^{\prime }},  \notag
\end{eqnarray}%
and%
\begin{eqnarray}
2\overline{\Gamma } &=&\left\langle \sigma _{B\overline{B}}v\right\rangle
sY_{B}+\left\langle \sigma _{BB}v\right\rangle sY_{\overline{B}}  \label{Gb}
\\
&&+\left\langle \sigma _{B\overline{B}}v\right\rangle ^{\prime }s^{\prime
}Y_{B^{\prime }}+\left\langle \sigma _{BB}v\right\rangle ^{\prime }s^{\prime
}Y_{\overline{B}^{\prime }},  \notag
\end{eqnarray}%
with $\sigma _{BB}=(1/2)\left( \sigma _{np}+\sigma _{nn}\right) $ and $%
\sigma _{B\overline{B}}=(1/2)\left( \sigma _{n\overline{p}}+\sigma _{n%
\overline{n}}\right) $.\footnote{%
Cross-sections for baryon interactions can be fitted using: $\sigma =\sigma
_{0}+\alpha c/v+\beta c^{2}/v^{2}$ with parameters obtains for literature 
\cite{cs1,cs2,cs3,cs4,cs5}, with: $\left\langle \sigma v\right\rangle =(4/%
\sqrt{\pi })c\sigma _{0}/\sqrt{x}+\alpha c+\left( \beta c/\sqrt{\pi }\right) 
\sqrt{x}.$} Equations (\ref{B1}) to (\ref{B4}) are stiff equations. They
have no analytical solutions, but they can be solved numerically by using a
linear multistep method based on the backward differentiation formula (BDF)
approach.\footnote{%
The ODE system under consideration is solved with a Python code using the
BDF mode of the function solve\_ivp of the SciPy module (https://scipy.org).}
The results of computations are shown and discussed in the next section.

\section{Results and discussion}

\label{results}

In the following, one sets $M_{B} = M_{P}$ following recent bounds \cite%
{Exp3,Exp4,cp1}.

\begin{figure}[ht!]
\centerline{\ \includegraphics[width=8.5cm]{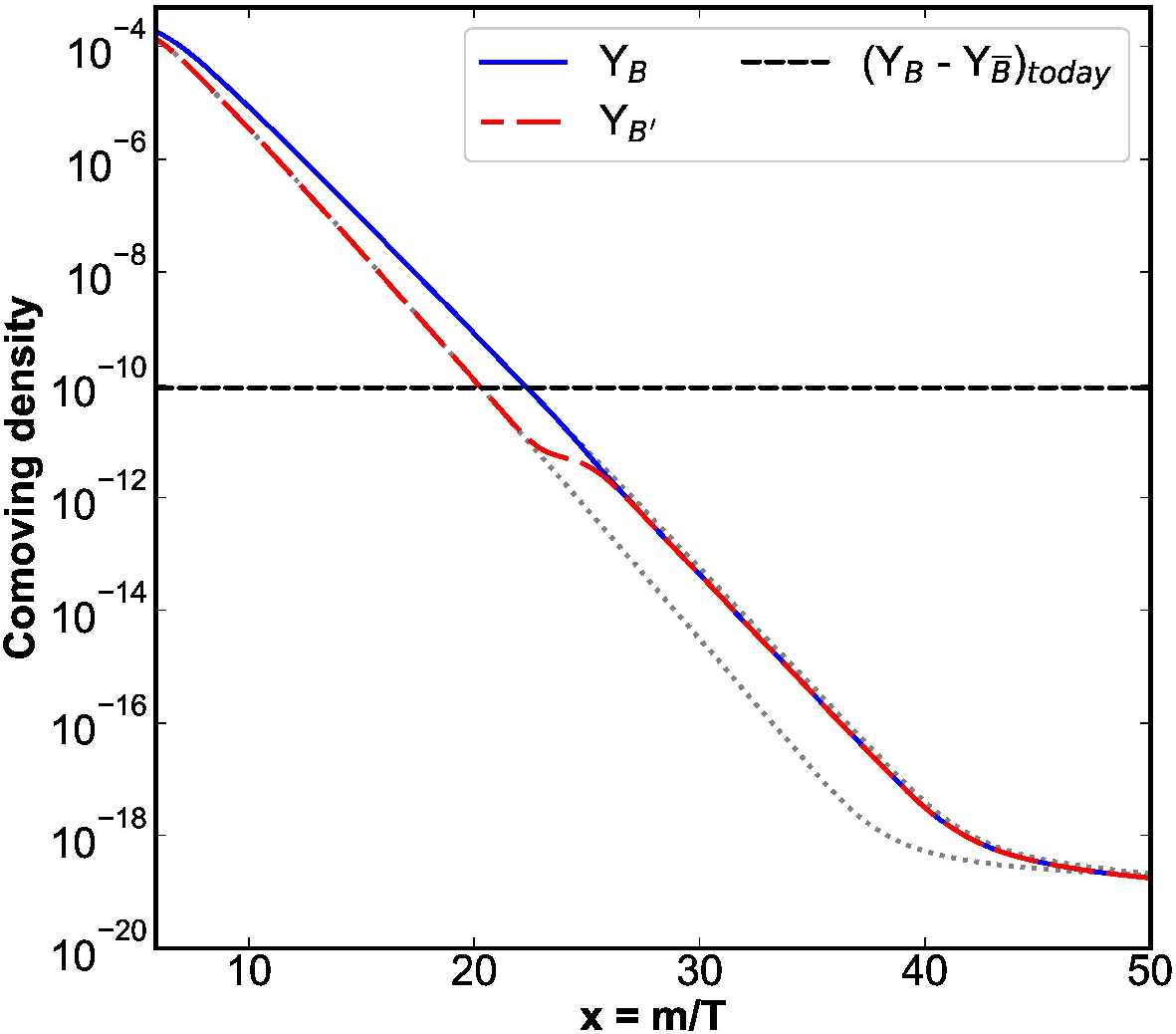}}
\caption{(Color online). Comoving densities $Y_{B}$ (superimposed with $Y_{%
\overline{B}}$) and $Y_{B^{\prime }}$ (superimposed with $Y_{\overline{B}%
^{\prime }}$) against $x$ for two coupled braneworlds but with no asymmetry (%
$\protect\delta =0$) and for $\protect\kappa =1.1$. Upper (respectively
lower) gray doted line corresponds to $Y_{B}$ and $Y_{\overline{B}}$
(respectively to $Y_{B^{\prime }}$ and $Y_{\overline{B}^{\prime }}$) when
branes are uncoupled. All the curves are superimposed when $\protect\kappa %
=1 $ and without coupling (not shown). Black dashed line is the current
asymmetry given by Eq. (\protect\ref{assy}).}
\label{fig:noCoupling+noAssym}
\end{figure}

Figure \ref{fig:noCoupling+noAssym} shows the behaviors of the comoving
densities $Y_{B}$, $Y_{\overline{B}}$, $Y_{B^{\prime }}$ and $Y_{\overline{B}%
^{\prime }}$ for $\kappa =1.1$ (i.e. $T^{\prime }$ is lower than $T$ by $9.1$%
\%), with coupling but without asymmetry ($\delta =0$). $Y_{B}$ and $Y_{%
\overline{B}}$ (respectively $Y_{B^{\prime }}$ and $Y_{\overline{B}^{\prime
}}$) in the visible brane (respectively in the hidden brane) are
indistinguishable. For the sake of comparison, one shows the comoving
densities for uncoupled branes (see caption), which are the expected
solutions of Eq.(\ref{BTE}). Although $Y_{B}$ and $Y_{B^{\prime }}$ (or $Y_{%
\overline{B}}$ and $Y_{\overline{B}^{\prime }}$) initially have different
dynamics due to different temperatures in each brane, when $x\approx 5$ all
the densities converge to share the same behavior. This describes the
thermalization of the two branes, which occurs due to their coupling through
neutron and antineutron exchanges. However, the lack of asymmetry (i.e. $%
\overline{\mathfrak{g}}=\mathfrak{g}=g$) cannot lead to baryogenesis.

\begin{figure}[ht!]
\centerline{\ \includegraphics[width=8.5cm]{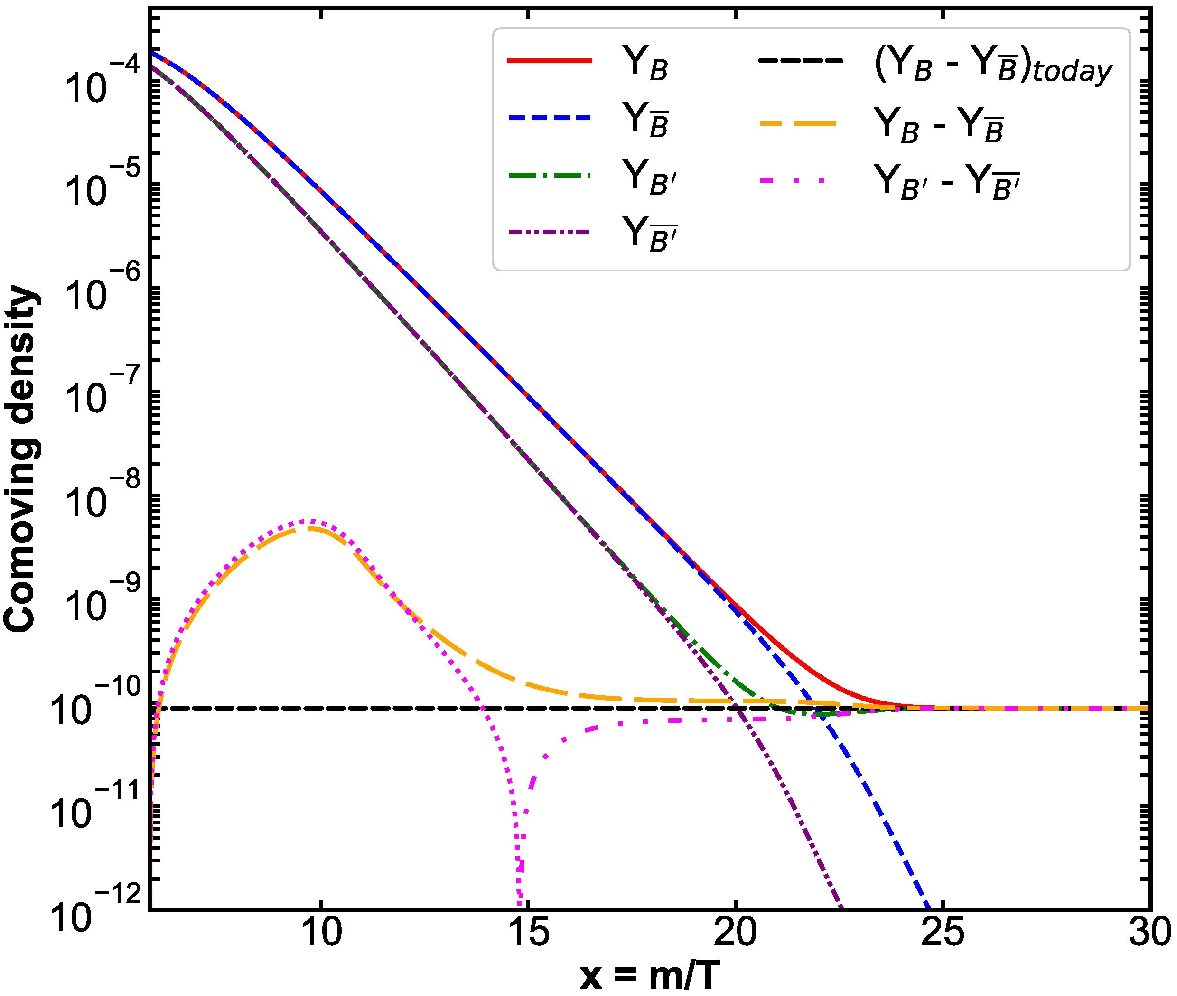}}
\caption{(Color online). Comoving densities $Y_{B}$, $Y_{\overline{B}}$, $%
Y_{B^{\prime }}$ and $Y_{\overline{B}^{\prime }}$ against $x$ with $\protect%
\kappa =1.1$, and a coupling between the two braneworlds with an asymmetry $%
\protect\delta =4.06\times 10^{-4}$. Orange dashed line is the difference
between populations of baryons and antibaryons. Pink line is the difference
between populations of hidden baryons and hidden antibaryons. The pink
dash-dot-dotted is for $Y_{B^{\prime }}-Y_{\overline{B}^{\prime }}>0$, while
the pink dotted line is for the opposite. Black dashed line is the current
asymmetry given by Eq. (\protect\ref{assy}).}
\label{fig:Assym}
\end{figure}

In figure \ref{fig:Assym}, all the Sakharov conditions are present: the
coupling between both branes leads to baryon number violation, the two
branes are not in thermal equilibrium (here $\kappa =1.1$), and an asymmetry
resulting in C/CP violation is introduced (in the present example $\delta
=4.06\times 10^{-4}$, see Eq. (\ref{asydef}) in section \ref{inter}). Such
conditions lead to baryogenesis and the current asymmetry between baryons
and antibaryons.

Figure \ref{fig:Assym} provides an explanation of the baryon-antibaryon
asymmetry mechanism. Early after QGPHG transition (before $x=10$), due to
C/CP violation, the swapping of antineutrons towards another brane is
enhanced compared to neutrons. Since the hidden brane has a lower
temperature than the visible brane, the net balance from the
matter-antimatter exchange between both branes promotes a decrease in
antineutrons in our brane and an increase in the hidden brane. As a result,
and due to the neutron-proton equilibrium (and the antineutron-antiproton
equilibrium) the antibaryon content decreases in our brane while the baryon
content tends to dominate (as shown by the orange dashed line). In contrast,
in the hidden brane the antibaryon content increases while the baryon
content tends to decrease (see pink dotted line).

In a late time after the QGPHG transition (after $x=10$), as soon as the
baryonic matter widely dominates the content of our visible brane, and due
to a higher temperature than in the hidden brane, baryons from our brane
feed the hidden brane, allowing for annihilation of antibaryons until the
matter-antimatter ratios reach the same values in both branes (pink
dash-dot-dotted and orange dashed line after $x=15$).

It should be noted that a positive asymmetry ($\delta >0$) favors a
two-brane Universe dominated by baryons, while an opposite asymmetry ($%
\delta <0$) leads to a Universe dominated by antibaryons in a comparable but
reversed proportion (not shown). Also, for $\kappa <1$, the roles of the
visible and hidden brane are simply reversed.

\begin{figure}[ht!]
\centerline{\ \includegraphics[width=8.5cm]{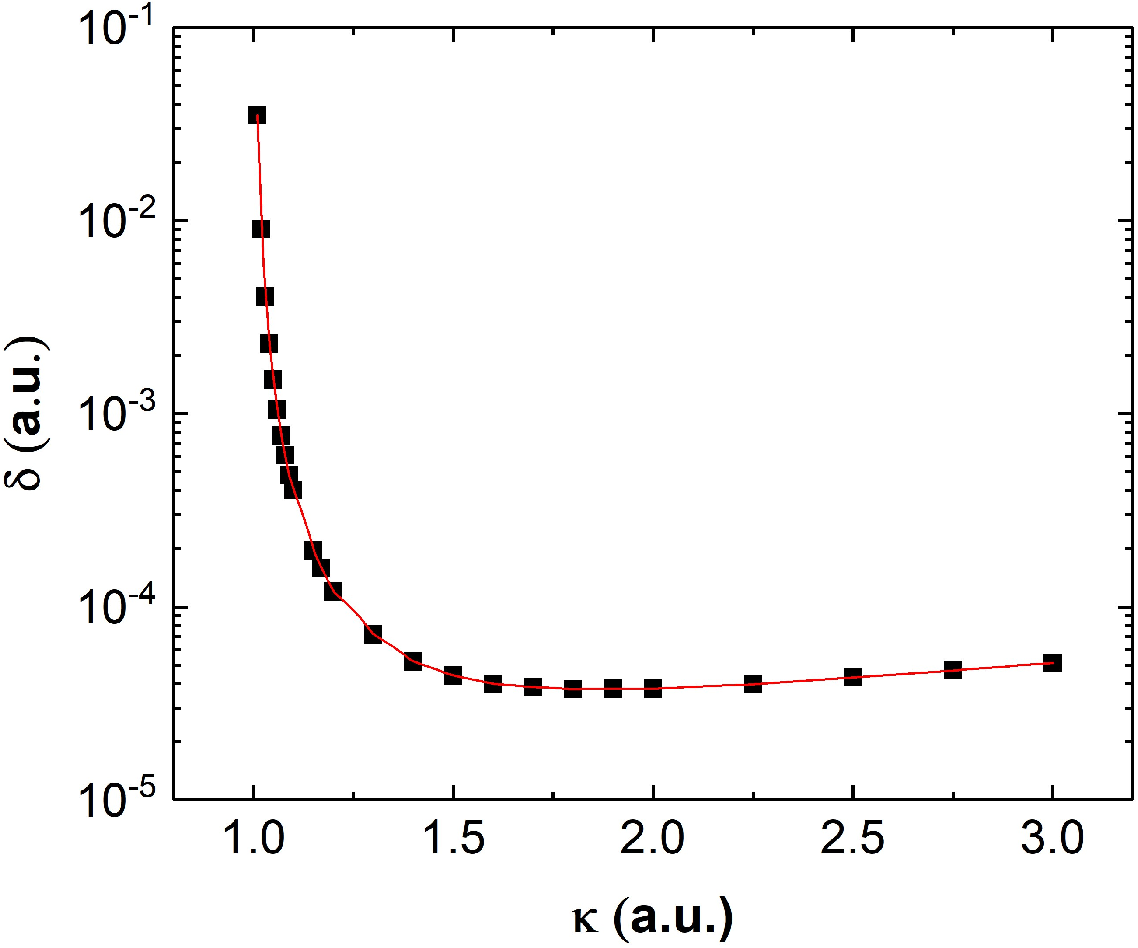}}
\caption{(Color online). Magnitude of the asymmetry $\protect\delta$ inducing the imbalance between baryons and
antibaryons observed today, against the ratio $\protect\kappa$ between the temperature in our visible braneworld
and the temperature in the hidden braneworld.}
\label{fig:Magn}
\end{figure}

Figure \ref{fig:Magn} shows the magnitude of C/CP-violation $\delta $ (see
Eq. (\ref{asydef}) in section \ref{inter}) against $\kappa $, for which one
gets the value of $Y_{B}-Y_{\overline{B}}$ observed today (see Eq. \ref{assy}%
) from computations. For $\kappa =1$ and $\kappa \gtrsim 3$, no value of $%
\delta $ can account for the observed imbalance between baryons and
antibaryons. However, a wide range of values for $\delta $ allows for the
imbalance of the baryon-antibaryon populations today observed as shown in
Fig. \ref{fig:Magn}. Thus, one gets (see Fig. \ref{fig:Magn}): 
\begin{equation}
4\times 10^{-5}<\delta <4\times 10^{-2}.  \label{ct}
\end{equation}%
These values have been reported on Fig. \ref{fig:f2}. The upper red dashed
line represents the upper limit $\delta =4\times 10^{-2}$ compatible with
the baryon-antibaryon imbalance, while the lower blue dashed line represents
the lower limit $\delta =4\times 10^{-5}$ allowing baryogenesis. As explain
previously in section \ref{inter}, Fig. \ref{fig:f2}, shows how the
magnitude of C/CP-violation $\delta $ depends on phase $\theta $ (see also
Eq. (\ref{var})), which is related to the electromagnetic fields in each
brane (see Eq. (\ref{ksi})). The values of $\theta $ that are compatible
with baryogenesis span a range of $177$ degrees. From a random point of
view, there is a very high probability -- almost a $1$ in $2$ chance -- that
the scalar field phase $\theta $ can promote baryogenesis. Moreover, from an
observational point of view, as the values of $Y_{B}-Y_{\overline{B}}$ must
fluctuate as $\theta $, then $Y_{B}-Y_{\overline{B}}$ must vary when the
primordial magnetic fields fluctuate following Eq. (\ref{ksi}). Subsequently, an 
important and challenging astrophysical endeavor would be the measurement of the baryon asymmetry, $Y_{B}-Y_{%
\overline{B}}$, across diverse areas of the observable universe. This data could then be associated with potential 
fluctuations in primordial magnetic fields to provide constraints on the current theoretical model. We do not develop this topic here as it is far beyond the
scope of the present paper, and we let it for future work.

The dynamics of leptogenesis is driven by baryogenesis in order to maintain
thermodynamic balance. As $Y_{p,eq}\approx Y_{n,eq}$, the neutron density
decreases due to matter exchange between branes, which causes the proton
population to also decrease in order to restore equilibrium. Therefore, $%
Y_{p}=Y_{n}$. This occurs through proton-electron capture, which is
thermodynamically favored. As a result, the electron density also decreases
while the neutrino density increases. One gets: $%
Y_{e-}=Y_{e-,eq}-(Y_{n,eq}-Y_{n})$ and $Y_{\nu }=Y_{\nu
,eq}+(Y_{n,eq}-Y_{n}) $. The same process occurs for antiparticles, but
antiproton-positron capture is favored. This causes the positron density to
decrease while the antineutrino density increases. One gets: $%
Y_{e^{+}}=Y_{e^{+},eq}-(Y_{\overline{n},eq}-Y_{\overline{n}})$ and $Y_{%
\overline{\nu }}=Y_{\overline{\nu },eq}+(Y_{\overline{n},eq}-Y_{\overline{n}%
})$.

By comparing the particle and antiparticle populations, one deduces: $%
Y_{e^{-}}-Y_{e^{+}}=(1/2)(Y_{B}-Y_{\overline{B}})$ and $Y_{\nu }-Y_{%
\overline{\nu }}=-(1/2)(Y_{B}-Y\overline{_{B}})$. This means that $Y_{L}-Y%
\overline{_{L}}=0$, i.e. the global leptonic number is zero. Furthermore,
positrons and antiprotons will be annihilated in such a way that each
remaining proton charge is compensated by an electron charge, thereby
maintaining the global neutrality of the Universe.

\section{Conclusion}

\label{conclusion}

Thanks to the low-energy limit of a two-brane universe -- resulting in a
noncommutative two-sheeted space-time -- it has been demonstrated that the
exchange of matter between the two branes does not occur at the same rate
for antimatter. This discrepancy arises from a violation of the C/CP
symmetry induced by a pseudo-scalar field that emerges due to the extension
of the electromagnetic gauge field in the two-brane system. This provides a
straightforward physical mechanism allowing baryogenesis to occur after the
quark-gluon era without stringent parameter constraints in cosmological
braneworld scenarios. Slight fluctuations of the baryon-antibaryon comoving
asymmetry, related to primordial magnetic fluctuations, could be a signature
of the model. To constrain the latter, it is suggested to attempt to measure 
$Y_{B}-Y_{\overline{B}}$ fluctuations in correlation with primordial
magnetic field fluctuations. Scenarios with definitions of the field
strength different from that used in the present paper could also be
explored in future work, both theoretically and experimentally. Ultimately, a 
thorough analysis of the dynamics involving additional particles -- such as 
other baryons, mesons, and leptons -- is planned in order to enrich the description of baryogenesis.

\section*{Acknowledgment}

The authors thank Patrick Peter for encouraging us to explore this topic as
well as for discussions and comments on an earlier draft of this paper.

\end{document}